\documentclass[aps,prb,amsmath,amssymb,amsfonts,nofootinbib,
superscriptaddress,twocolumn]{revtex4}
\usepackage{graphicx}
\usepackage{dcolumn}
\usepackage{bm,bbm}
\usepackage{amsmath}
\usepackage{amssymb}
\usepackage{color}

\usepackage{latexsym,multirow,hyperref}
\usepackage{graphicx}
\usepackage{times,psfrag,subfigure}
\usepackage{amsmath}
\usepackage{dcolumn}
\usepackage{color}
\usepackage{soul}
\usepackage{latexsym,amsmath,amssymb,bm,euscript}
\usepackage{threeparttable}
\usepackage{hyperref}

\newcommand{\beq}{\begin{equation}}
\newcommand{\eeq}{\end{equation}}
\newcommand{\beqarray}{\begin{eqnarray}}
\newcommand{\eeqarray}{\end{eqnarray}}

\def\cH{{\cal H}}

\newcommand{\ket}[1]{| #1 \rangle}

\newcommand{\bee}{\begin{equation}}
\newcommand{\ee}{\end{equation}}
\newcommand{\bma}{\begin{pmatrix}}
\newcommand{\ema}{\end{pmatrix}}
\newcommand{\balig}{\begin{align}}
\newcommand{\ealig}{\end{align}}

\newcommand{\bI}{\mathbbm{1}}
\newcommand{\ba}{\begin{eqnarray}}
\newcommand{\ea}{\end{eqnarray}}

\newcommand{\ignore}[1]{}

\newcommand{\cP}{\mathcal{P}}
\newcommand{\bR}{\bf R}
\newcommand{\br}{\bf r}
\newcommand{\cR}{\mathcal{R}}

\begin{document}

\title{Interaction-enabled topological phases in topological insulator -superconductor heterostructures}
\author{D. I. Pikulin}
\email{pikulin@phas.ubc.ca}
\affiliation{Department of Physics and Astronomy and Quantum Matter Institute, University of British Columbia, Vancouver, BC, Canada V6T 1Z1} 
\author{Ching-Kai Chiu}
\affiliation{Department of Physics and Astronomy and Quantum Matter Institute, University of British Columbia, Vancouver, BC, Canada V6T 1Z1}
\author{Xiaoyu Zhu}
\affiliation{Department of Physics and Astronomy and Quantum Matter Institute, University of British Columbia, Vancouver, BC, Canada V6T 1Z1}
\affiliation{National Laboratory of Solid State Microstructures and Department of Physics, Nanjing University - Nanjing 210093, China}
\author{M. Franz}
\affiliation{Department of Physics and Astronomy and Quantum Matter Institute, University of British Columbia, Vancouver, BC, Canada V6T 1Z1}

\begin{abstract}
 Topological phases of matter that depend for their existence on interactions are fundamentally interesting and potentially useful as platforms for future quantum computers. Despite the multitude of theoretical proposals the only interaction-enabled topological phase experimentally observed is the fractional quantum Hall liquid. To help identify other systems that can give rise to such phases we present in this work a detailed study of the effect of interactions on Majorana zero modes bound to vortices  in a superconducting surface of a 3D topological insulator. This system is of interest because, as was recently pointed out, it can be tuned into the regime of strong interactions. We start with a 0D system suggesting an experimental realization of the interaction-induced $\mathbb{Z}_8$ ground state periodicity previously discussed by Fidkowski and Kitaev. We argue that the periodicity is experimentally observable using a tunnel probe. We then focus on interaction-enabled crystalline topological  phases that can be built with the Majoranas in a vortex lattice in higher dimensions. In 1D we identify an interesting exactly solvable model which is related to a previously discussed one that exhibits an interaction-enabled 
topological phase. We study these models using analytical techniques, exact numerical diagonalization (ED) and density matrix renormalization group (DMRG). Our results confirm the existence of the interaction-enabled topological phase and clarify the nature of the quantum phase transition that leads to it.  We finish with a discussion of models in dimensions 2 and 3 that produce similar interaction-enabled topological phases.
\end{abstract}
\maketitle

\section{Introduction}
Non-interacting electron systems have recently provided a new playground for experimental and theoretical research -- topological insulators (TIs) and topological superconductors (TSCs).\cite{Has10, Qi11, Fra13} The hallmarks of these materials are symmetry-protected gapless edge modes. Both fundamental and crystalline symmetries can play a role in protecting these states.\cite{LFu11, And13, Chi15} The TIs and TSCs are extensively studied both on their own and as building blocks in engineered exotic quantum states.\cite{Ali12, Bee13, Sta13, Ell15, Cla13, Lin12, Vae13, Bar12}

Majorana bound state is an example of such an edge state, which can emerge both at the edge of a 1D TSC and in the vortex of 2D TSC, which may be either intrinsic or realized at the interface between a 3D TI and an ordinary superconductor.\cite{Ali12, Bee13, Sta13, Ell15, Fu_proximity} Such Majorana bound states occur at zero energy and are thus often called Majorana zero modes (MZMs). The effort to detect such MZMs has been focused on the zero-bias conductance peak,\cite{Mou12, Das12, Den12} indicative of the density of states at zero energy, which is, however, not a definitive proof of a Majorana bound state.\cite{Pik12} The physics of the MZMs still to be observed is much richer: most importantly, upon adiabatic exchange they behave as non-abelian anyons and provide a pathway to (non-universal) topological quantum computation.\cite{Nay08} 

It is known that the electron-electron interactions enrich the palette of topological states. The first discovered topological material with inherently strong electron-electron interactions is the fractional quantum Hall effect.\cite{Tsu82, Lau83} For certain filling fractions the emergent quasiparticles are predicted to be Majoranas, or even more exotic Fibonacci anyons, that may provide a platform for universal quantum computation.\cite{Nay08,Jai07} These systems require extreme conditions to be probed and used: very clean samples, low temperature, and high magnetic fields. One way to avoid these complications is to engineer the Majorana bound states\cite{Lut10, Ore10} and use the interactions between them to construct universal quantum computer. Majorana surface codes realizing exotic non-abelian quasiparticles allowing for universal quantum computation, have been theoretically proposed. \cite{Vij15}

For the experimental realization of these proposals there are several preliminary steps to be made. Though there have been indications of the Majorana bound states in the conductance measurements in several systems, the possibility to induce interactions between them still remains unconfirmed. In the present paper we endeavor to fill this gap from the theory side by suggesting realistic setups that can be used to probe the interactions between Majoranas.

As a basis we use a system proposed recently as a convenient platform to study strong interactions between Majorana zero modes: vortices in a surface of 3D TI with induced superconductivity and the chemical potential tuned to the Dirac point. It was argued that for this special point direct Majorana-Majorana tunneling is absent due to the additional chiral symmetry of the underlying physical system.\cite{teo1,galitsky2,Tew12,Chi15a} The system of MZMs  is then in the so-called symmetry class BDI and can be viewed as respecting a fictitious time-reversal symmetry\cite{Chi15b} $\bar{\Theta}$ such that $\bar{\Theta}^2=1$. The dominant term in the low-energy Hamiltonian is the 4-Majorana interaction that arises from the Coulomb interaction between the constituent electrons.\cite{Chi15a}  A number of interesting phases have been identified in this setup\cite{lahtinen1,lahtinen2,tliu1,vafek1} including some interaction-enabled phases.\cite{Chi15b, Rah15, Cobanera15, Rah15a}

In the present paper we study the effects of interactions on the phases of the MZMs in dimensions 0-3. We consider systems with time-reversal symmetry $\bar{\Theta}$, and the main focus of the paper is on 0D and 1D. We start with the 0D case. We first argue for $\mathbb{Z}_8$ Fidkowski-Kitaev periodicity of the ground state degeneracy \cite{Fid10, Fid11} to coincide with the degeneracy of the entanglement spectrum.\cite{turner11} The problem can be stated as follows. Given the set of $n$ MZMs such that no direct tunneling between them is allowed, what is the ground state degeneracy in the presence of {\em generic} 4-fermion interactions? By a combination of simple arguments and direct computations   we determine the degeneracy for $n=1\dots 8$ and then present an argument that the pattern repeats itself with the periodicity of 8.
We then suggest a tunneling probe as a way to experimentally observe the expected $\mathbb{Z}_8$ periodicity. The probe would reveal the absence of the zero-bias peak for the total vorticity being a multiple of $4$, in contrast with the non-interacting case for which we expect the zero bias peak to be absent for even total vorticity.

The second part of the paper focuses on the interaction-enabled crystalline topological phases in dimensions 1, 2 and 3.  As previously noted by Lapa, Teo and Hughes (LTH),\cite{Lap14} in the presence of inversion and the fictitcious time-reversal symmetry there exist no topological phases in non-interacting 1D models. This is because the inversion symmetry maps the integer topological invariant $\nu$ to $-\nu$. For an inversion-symmetric system, therefore $\nu=-\nu$, implying $\nu=0$ as the only solution. In the presence of interactions the integer classification changes to $\mathbb{Z}_8$. Under $\mathbb{Z}_8$, remarkably, the equation $\nu=-\nu$ has a non-trivial solution $\nu=4$, indicating a genuine interaction-enabled crystalline topological phase. We discuss a physical
setup that realizes such a phase in the system of vortices and antivortices in the surface of a 3D TI. We start this discussion by introducing a non-trivial  but exactly solvable ``2-leg LTH model'' that is closely related to the LTH model but does not exhibit the topological phase. It has two distinct phases, distinguished by a conventional broken symmetry. The 2-leg LTH model can be mapped onto a collection of the Kitaev chain models, which allows for exact solution. We use this model to benchmark our numerical simulations which we then employ to study the ``4-leg LTH model'' which has an interaction-enabled topological phase. The 4-leg model is exactly solvable only in two extreme limits, the strong coupling limit shows the topological phase. We use ED and DMRG techniques to study its ground state degeneracy, central charge and the entanglement spectrum away from the exactly solvable limits. We show that the topological phase extends over a significant part of the phase diagram. 

We close by proposing variants of the interaction-enabled topological crystalline phases in 2D and 3D that can be built, at least in principle, from the ingredients at hand. As in 1D, these phases require interactions for their existence and exhibit anomalous edge or surface states protected by a combination of time-reversal and crystalline symmetry such as inversion or discrete rotation.  

\section{Physical  system}\label{FuKane}

In this section we introduce our platform for strongly interacting Majorana fermions and review some facts that will be important in what follows. We consider vortices in the surface of a strong topological insulator in the proximity with an $s$-wave superconductor, which was first proposed by Fu and Kane.\cite{Fu_proximity} When the chemical potential is tuned to the Dirac point a fictitious  time reversal symmetry $\bar{\Theta}$ emerges. Hopping of MZMs located in the individual vortices is forbidden by $\bar{\Theta}$  and the second major effect, interaction, becomes dominant.\cite{Chi15a, Chi15b}
Here we will only review the surface physics  briefly and point out the important for the present work properties of the Hamiltonian and the wavefunctions in the Fu-Kane model. The Hamiltonian of such a system in the Nambu four-vector basis $(c_{\uparrow \br }, c_{\downarrow \br }, c^\dagger_{\downarrow \br}, -c^\dagger_{\uparrow \br})$ reads
\begin{align}
H_{\rm{FK}}(\mathbf{r},\mathbf{p}) = &v p_x \sigma_x \tau_z + v p_y \sigma_y \tau_z - \mu \tau_z \nonumber \\ &+ \mathrm{Re}\Delta(\mathbf{r})\tau_x + \mathrm{Im}\Delta(\mathbf{r}) \tau_y, \label{eq:Hamiltonian}
\end{align}
where $\sigma$ and $\tau$ are Pauli matrices in the spin and Nambu basis respectively. Particle-hole  symmetry automatically is preserved with $\Xi=  \tau^y\sigma^y  K$,  where $K$ represents complex conjugation. In the absence of the vortices the system also preserves physical time-reversal symmetry, $\Theta = \sigma_y K$. By \emph{fine tuning} of the system, namely adjusting $\mu$ to zero, the system respects time-reversal symmetry with $\bar{\Theta}=\tau^x\sigma^x  K$. Notice that the physical time-reversal symmetry, which is broken in the presence of magnetic field, squares to $-1$, as it should  for spin-${1\over 2}$ particles, but the $\bar{\Theta}^2=1$, which is allowed for an \textit{accidental} symmetry. Furthermore, chiral symmetry is automatically preserved with the  operator $\Pi=\Xi\bar{\Theta}=-\tau^z\sigma^z$. In the following we focus on the $\mu=0$ situation with the extra symmetry. 

If we assume a vortex at the origin, the gap profile in polar coordinates is $\Delta(\mathbf{r}) = \Delta(r) e^{i n \phi}$, where $n$ is the vorticity of the vortex. For a system respecting $\bar{\Theta}$ and $\Xi$ the vorticity is the same as the total number of the MZMs inside the vortex as we will see shortly. We call the cases with $n>0$ vortices, and $n<0$ anti-vortices. The wavefunctions for these two cases are respectively
\begin{align}
\Phi_\downarrow=(0,u_\downarrow,u_\downarrow^*,0)^T,\quad \Phi_\uparrow=(u_\uparrow,0,0,-u_\uparrow^*)^T \label{vortex antivortex}
\end{align}
These wavefunctions belong to the opposite eigenvalues $\pm 1$ of the chirality operator $\Pi$. Importantly, the symmetry $\Pi$ forbids the hybridization between two Majorana zero modes with the same chirality.\cite{teo1,galitsky2,Tew12,Chi15a}.

To simplify the Hamiltonian and bring it into block off-diagonal form we perform a unitary transformation $H_{\rm{FK}}\to U H_{\rm{FK}} U^{-1}$ with
\begin{align}
U = \begin{pmatrix}
1 & 0 & 0 & 0 \\
0 & 0 & 0 & 1 \\
0 & 0 & 1 & 0 \\
0 & 1 & 0 & 0
\end{pmatrix}.
\end{align}
This renders the equations for the zero-mode for both vorticities
\begin{align}
D(\mathbf{r}) \chi_\downarrow(\mathbf{r}) = 0,\quad D^\dag(\mathbf{r}) \chi_\uparrow = 0,\label{eq:D}
\end{align}
where
\begin{align}
D = \begin{pmatrix}
e^{-i n \varphi - i \alpha} \Delta_0(r) & e^{-i \phi}\left( - i \partial_r - \frac{\partial_\varphi}{r}\right)\\
- e^{i \phi}\left( - i \partial_r + \frac{\partial_\varphi}{r}\right) & e^{i n \varphi + i \alpha} \Delta_0(r)
\end{pmatrix}
\end{align}
and $\Phi_\downarrow=U(0,\chi_\downarrow)^T$, $\Phi_\uparrow=U(\chi_\uparrow,0)^T$.
We thus establish the basic building block of the setups we will consider -- a vortex of vorticity $n$, which carries $n$ Majorana zero modes. If $n=1$, then the zero mode obtained from the equation \eqref{eq:D} is automatically the Majorana zero mode, otherwise MZMs can be obtained as linear combinations of these modes. 

In what follows we shall call $\alpha_j$ the creation operators of the MZMs from the $\downarrow$ sector and $\beta_j$ from the $\uparrow$ sector. We see from the explicit form of the Majorana wavefunctions in the two spin sectors, see Eq.\ \eqref{vortex antivortex}, that the two types of MZMs transform differently under the emergent time reversal, namely \cite{Chi15b}
\begin{equation}\label{op1}
\bar{\Theta} \alpha_j\bar{\Theta}^{-1}=\alpha_j, \ \ \ \ 
\bar{\Theta} \beta_j\bar{\Theta}^{-1}=-\beta_j.
\end{equation}
Along with the property $\bar{\Theta} i\bar{\Theta}^{-1}=-i$ this implies that one can construct a complex fermion $c_j={1\over 2}(\alpha_j+i\beta_j)$ that transforms naturally \cite{Fid10, Fid11,Lap14} under $\bar{\Theta}$, as $\bar{\Theta} c_j\bar{\Theta}^{-1}=c_j$. In the rest of this paper we shall study systems of MZMs described by $\bar{\Theta}$-invariant Hamiltonians that contain both 2-fermion ``tunneling'' terms and 4-fermion interaction terms. The former arise from the wavefunction overlaps between MZMs while the latter originate from the interactions between the constituent electron degrees of freedom. These include the direct Coulomb repulsion, as well as interactions mediated by other degrees of freedom such as phonons, which may be attractive. 

In a $\bar{\Theta}$-invariant Hamiltonian Eq.\ (\ref{op1}) allows tunneling terms between different types of MZMs of the form  
\begin{equation}\label{op2}
\cH_{\rm kin}=i\sum t_{ij} \alpha_i\beta_j,
\end{equation}
but prohibits those between the same chirality such as $i \alpha_i \alpha_j$  and $i \beta_i \beta_j$. Interaction terms with an even number of $\alpha$ operators such as $\alpha_i \alpha_j \alpha_k \alpha_m$, $\beta_i \beta_j \beta_k \beta_m$ and $\alpha_i \alpha_j \beta_k \beta_m$ are allowed but those with an odd number are prohibited. An interesting consequence of this structure arises in a system that is composed solely of vortices (and no antivortices). According to the above discussion all tunneling terms are then forbidden and the kinetic energy is quenched.  Interaction terms however are allowed by symmetry and the system is, therefore, inherently strongly interacting. In a realistic setting, the chemical potential will always be slightly detuned from zero (either globally or locally due to disorder) leading to non-zero $\cH_{\rm kin}$. However, as long as this detuning is small compared to the interaction scale $g$ the system must be viewed as strongly interacting. 

In Sec.\ III below we work with interactions only (assuming $\mu=0$ and the presence of only vortices in the system) while in Sec.\ IV we consider interacting systems composed of both vortices and antivortices that exhibit interactions as well as hopping. In this way we establish experimentally realizable strongly interacting phases of the Majorana bound states.

\section{Fidkowski-Kitaev periodicity}

In their seminal work Fidkowski and Kitaev\cite{Fid10, Fid11} predicted that in the presence of interactions the $\mathbb{Z}$ topological classification of a 1D systems of fermions with symmetry $\bar{\Theta}$ (class BDI) breaks down to a periodic $\mathbb{Z}_8$ one. That is, the gapped phase with the winding number $\nu=8$ can be continuously deformed to the trivial phase by turning on local interactions without any gap closing. The original non-trivial phase possesses $\nu$ Majorana zero modes at each edge. The above result fundamentally depends on the fact that symmetry-preserving interactions can completely remove the $2^4$-fold ground state degeneracy of $8$ MZMs but {\em cannot} do it for any smaller number of MZMs.  Fidkowski and Kitaev considered a very specific, highly symmetric, form of interactions to demonstrate the above effect. How general or finely tuned  the interactions need to be to produce the  unique ground state for 8 MZMs however remained unclear. Here we show that under quite general conditions the ground state degeneracy of $8$ Majoranas is split by Coulomb interactions between the constituent electrons. Moreover, we give a general argument showing the $\mathbb{Z}_8$ periodicity of the ground state degeneracy in a system of $n$ Majoranas in accordance with the Fidkowski-Kitaev result. We also explain how the resulting $\mathbb{Z}_8$ periodicity induced by interactions can be experimentally observed.
\begin{figure}[t]
\includegraphics[width = \linewidth]{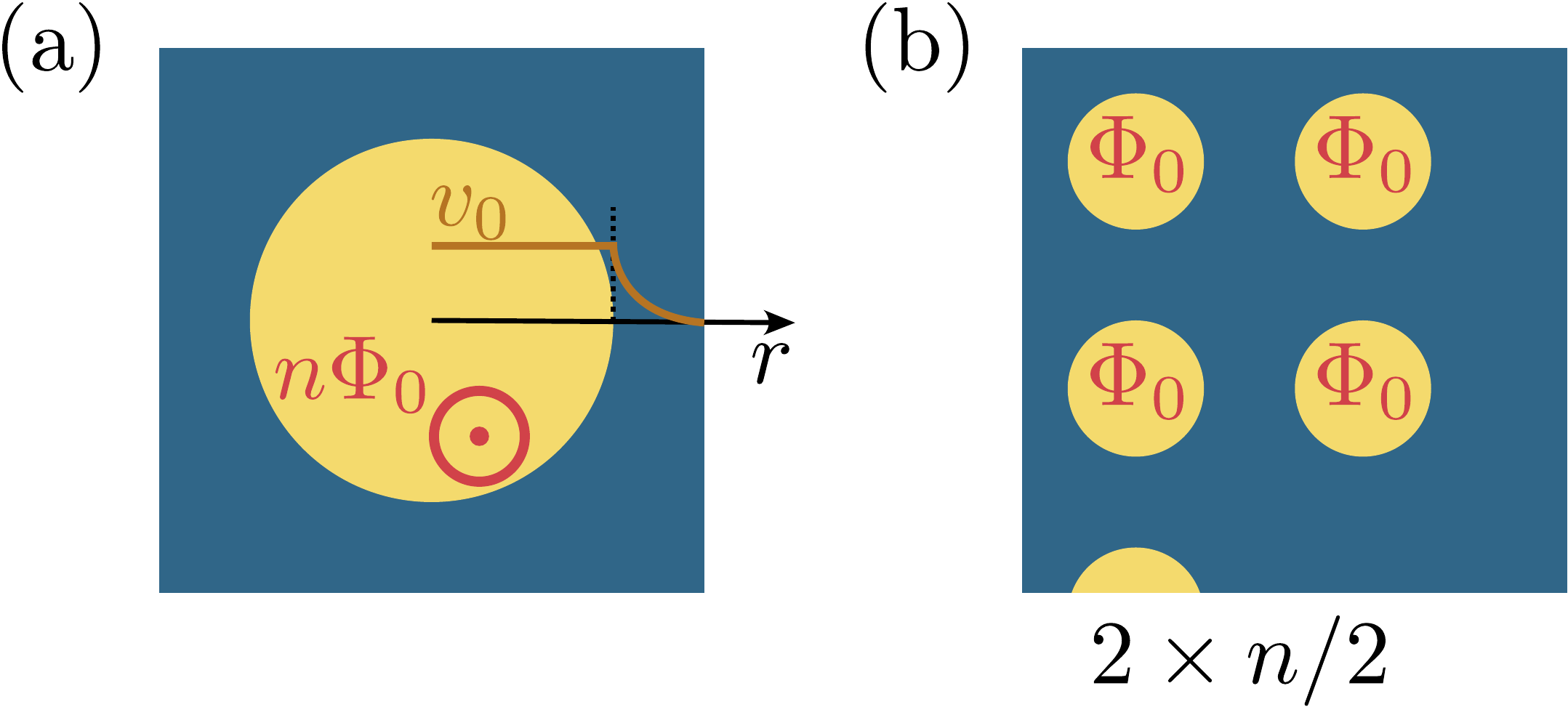}
\caption{Top view on the setups we suggest to observe the $\mathbb{Z}_8$ periodicity of the BDI Majorana model. Blue is the area covered by the superconductor, and yellow is the bare surface of the topological insulator. a) A single large hole in the superconductor. Similar to Corbino disk geometry all the magnetic field goes through the hole. Total flux is $n\Phi_0$ and the total number of MZMs in the non-interacting model is $n$. b) The lattice of vortices. We consider a square lattice with $2\times n/2$ vortices. If $n$ is odd, there are $(n-1)/2$ lines of $2$ Majoranas and one line with just one, similar to the $n=5$ case in the figure. Such an arrangement can be created by introducing strong pinning centers into the superconductor.
}\label{fig:setups_nMajoranas}
\end{figure}

In this section we are interested in compact 0D structures with the total vorticity $n$. We assume that $\mu=0$ and no tunneling terms between MZMs are therefore allowed although generic 4-fermion interaction terms are present. We consider two specific physical setups realizing such a system, shown in Fig. \ref{fig:setups_nMajoranas}. The first  is a superconductor with a large hole put on top of the TI with $n$ superconducting flux quanta threaded through the hole as outlined in Fig.\ \ref{fig:setups_nMajoranas}(a). This geometry can be viewed as a Corbino disk familiar from many superconducting applications. We therefore expect that this configuration is the easiest for experimental access. The second setup  is an array of single vortices illustrated in Fig. \ref{fig:setups_nMajoranas}(b). This requires more fine-tuning, for example, creating strong pinning centers for the vortices, but is ultimately also accessible experimentally, especially since our qualitative results do not depend on the exact geometric arrangement of vortices. We discuss the tunneling conductance signature of the interacting system and how it is modified as compared to the non-interacting case.

A detailed calculation of the energy spectra in these two systems shows that the same ground state periodicity of 8 in $n$ is observed in both. We generalize this result by providing an argument which proves the connection between the periodicity of the entanglement spectrum and the periodicity of the ground state degeneracy for a system of $n$ Majoranas with generic 4-fermion interactions.
We also comment on the applicability of this argument to the physical realizations of the interacting Majorana model suggested above.

\subsection{Giant  vortex in a Corbino geometry}

We start our discussion by considering the setup depicted in Fig. \ref{fig:setups_nMajoranas}(a). We envision a hole in the superconducting coating, through which a number $n$ of superconducting flux quanta $\Phi_0=hc/2e$  is threaded. This is equivalent to systems already studied in the context of TIs.\cite{Soc15} In this Section we will consider one sign of vorticity, positive for definiteness. We search for the solution of Eq.\ (\ref{eq:D}) in the form 
\begin{align}
\chi_m(\mathbf{r}) = \frac{1}{\sqrt{2}}\begin{pmatrix}
e^{i((n - 1 - m)\varphi + \alpha/2 - \pi/4)} u_m(r)\\
e^{-i(m\varphi + \alpha/2 - \pi/4)} v_m(r)
\end{pmatrix},
\end{align}
to obtain
\begin{align}
\left\{\begin{array}{c}
\Delta_0(r) u_m(r) + \left(\partial_r - \frac{m}{r}\right) v_m(r) = 0,\\
\Delta_0(r) v_m(r) + \left(\partial_r - \frac{n - 1 - m}{r}\right) u_m(r) = 0.
\end{array}\right. \label{eq:J-R}
\end{align}

Let us now take into account the practical realization of this giant vortex. We imagine a hole drilled in the superconductor, thus leaving the order parameter zero inside the hole and large outside, see Fig. \ref{fig:setups_nMajoranas}(a). Thus inside the hole the Eq.\ \eqref{eq:J-R} reduces to
\begin{align}
\left\{\begin{array}{c}
\left(\partial_r - \frac{m}{r}\right) v_m(r) = 0,\\
\left(\partial_r - \frac{n - 1 - m}{r}\right) u_m(r) = 0,
\end{array}\right.
\end{align}
having the obvious solutions
\begin{eqnarray}
v_m(r) &\propto& r^m, \\
u_m(r) &\propto& r^{n-1-m}.
\end{eqnarray}
Notice that when either of the powers is negative, the solution becomes non-normalizable at $r=0$, similar to the arbitrary gap profile situation discussed first by Jackiw and Rossi.\cite{Jac81}

The boundary condition at the external radius $r_0$ of the hole is determined by the absence of solutions growing into the superconductor. The equation \eqref{eq:J-R} under the assumption of large gap inside the superconductor gets simplified to
\begin{align}
\left\{\begin{array}{c}
\Delta_0(r) u_m(r) + \partial_r  v_m(r) = 0,\\
\Delta_0(r) v_m(r) + \partial_r u_m(r) = 0.
\end{array}\right.
\end{align}
This requires
\begin{align}
u_m(r_0) = v_m(r_0).
\end{align}
Therefore inside the hole we have
\begin{eqnarray}
v_m(r) &=& A (r/r_0)^m, \\
u_m(r) &=& A (r/r_0)^{n-1-m}.
\end{eqnarray}
Normalization of the wavefunction then requires
\begin{align}
\int_0^{r_0} r dr( v_m^2(r) + u_m^2(r))  = 1,
\end{align}
which finally gives
\begin{align}
A = r_0^{-2} \left[\frac{1}{2m + 1} + \frac{1}{2n - 2m - 1}\right]^{- 1/2}.
\end{align}

We have thus found the wavefunctions of the Andreev bound states in the hole at zero energy. To proceed in a uniform fashion we change to the Majoranas basis.
For that we first rewrite the solutions of the BdG equation in the second quantization form
\begin{align}
\chi_m = \frac{1}{\sqrt{2}} \int d\mathbf{r}\left(e^{i((n-1-m)\varphi + \alpha/2)} u_m(r) c_{\downarrow\br}^\dag\right.\nonumber \\ \left.+ e^{-i(m\varphi + \alpha/2)} v_m(r) c_{\downarrow\br}\right).
\end{align}

We notice that the connection between Andreev bound state with indices $m$ and $n-1 - m$: $\chi_m = \chi_{n - 1- m}^\dag$ as $v_{n-1-m}=u_m$. Thus  for $m<(n-1)/2$ we can define MZM operator as
\begin{equation}
\alpha_m= \frac{\chi_m + \chi_{n-1-m}}{\sqrt{2}} 
\end{equation}
and for $m>(n-1)/2$
\begin{equation}
\alpha_m = i \frac{\chi_m - \chi_{n-1-m}}{\sqrt{2}} 
\end{equation}
For $m=(n-1)/2$ and $n$ odd we also have
\begin{align}
\alpha_m = \chi_m.
\end{align}

The interaction between MZMs arises from the underlying electron-electron interaction 
\begin{align}
\cH_{\rm int}=\frac{1}{2}\int d\mathbf{r}d\mathbf{r}'\rho(\mathbf{r}) V(\mathbf{r}, \mathbf{r}') \rho(\mathbf{r}'),\label{eq:matrix_element}
\end{align}
where $\rho(\mathbf{r})=\sum_\sigma c_{\sigma\br}^\dagger c_{\sigma\br}$ represents the electron density operator. 
The interaction matrix element between MZMs can be computed by expressing the charge density operators in terms of the eigenstates of $H_{\rm FK}$ and projecting onto the zero-energy Majorana subspace, as described in Ref.\ \onlinecite{Chi15a}. One then obtains MZM interaction Hamiltonian of the form 
\begin{align}\label{hint}
\cH_{\rm int}=\sum_{ijkl}g_{ijkl}\alpha_i \alpha_j \alpha_k \alpha_l.
\end{align}
The matrix element $g_{ijkl}$ is computed numerically from the eigenfunctions $u_m$, $v_m$ and the electron-electron interaction potential $V$. 
In what follows we will use screened Coulomb potential
\begin{align}
V(\mathbf{r}, \mathbf{r}') = \frac{V_0}{|\mathbf{r} - \mathbf{r}'|} 
e^{ - |\mathbf{r} - \mathbf{r}'| / r_s}
\end{align}
where $V_0$ is the Coulomb interaction strength, and $r_s$ is the screening length. We use $r_s\to\infty$ below, but we have checked that the finite screening length does not change the results qualitatively. We will work in the units where the radius of the hole is $1$. In these units if the hole has a radius of $100$~nm and taking into account the dielectric constants of the TI materials, ranging from  $\epsilon=20$ for HgTe to $\sim 200$ for $\mathrm{Bi}_2\mathrm{Te}_3$,\cite{But10} interaction strength is $V_0 \approx 0.72-0.072$~meV, and the magnetic field to create vorticity $8$ in the core is $\approx 65$~mT.

\begin{figure}[t]
\includegraphics[width = 0.8\linewidth]{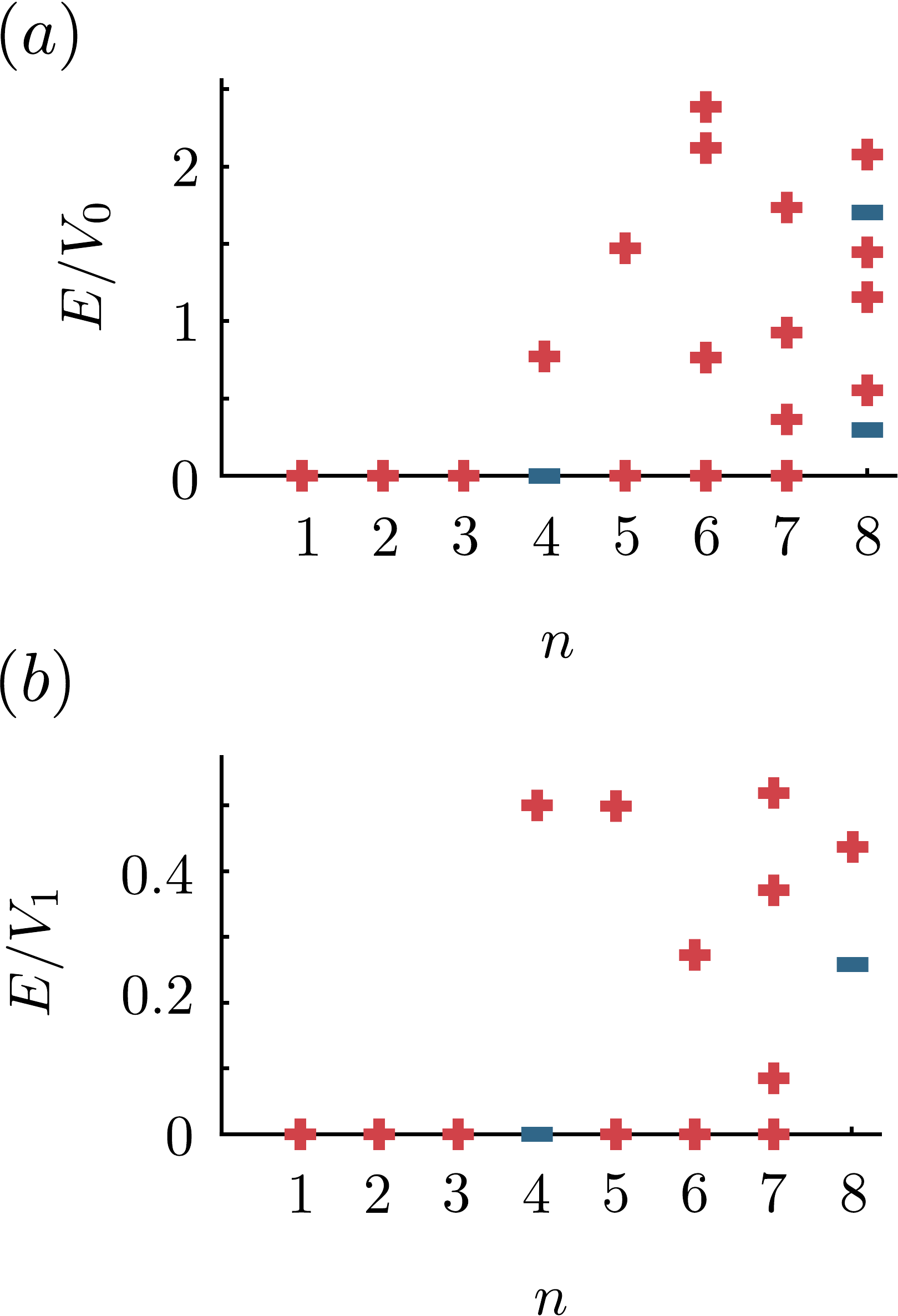}
\caption{Positions of energy levels and conductance peaks as a function of the total vorticity $n$  for the two setups depicted in Fig.\ \ref{fig:setups_nMajoranas} (a) and (b) correspondingly. Red cross depicts the position of the energy level observable as Lorenzian peaks in tunneling conductance. Blue bar shows the energy level different from the ground state by an even number of fermions. Such level is not observable by single-electron tunneling.
}\label{fig:setups_conductance}
\end{figure}

By exact numerical diagonalization of the manybody Hamiltonian $\cH_{\rm int}$ with all 4-fermion interactions as computed above we find the energy spectra of the system for $n$ up to $16$. The low-lying eigenenergies for $n$ up to $8$ are displayed in Fig.\ \ref{fig:setups_conductance}(a). The structure here is easy to understand for small $n$. For $n=1,2,3$ no 4-fermion term can be constructed, $\cH_{\rm int} =0$, and all the states are at zero energy. For $n=4$ we have $\cH_{\rm int}=g\alpha_1\alpha_2\alpha_3\alpha_4$ which results in a 2-fold degenerate ground state and 2-fold degenerate excited state. For $n>4$ progressively larger number of interaction terms can be constructed and the energy levels must be found numerically. From this we deduce the ground state degeneracies and find  that they repeat in $n$ with periodicity of  $8$ as predicted by Fidkowski and Kitaev \cite{Fid10, Fid11}. The degeneracies are listed in Table \ref{tab:GS_degeneracy} and agree with degeneracies of the entanglement spectrum for the corresponding topological phases predicted in Ref.\ \onlinecite{turner11}. This is true for a range of parameters, in particular we verified this result for different screening lengths $r_s$. Based on this observation we hypothesize that the periodicity of the ground state degeneracies holds for larger vorticities and for a generic arrangement of vortices. We will prove this statement below and specify more carefully the conditions under which it is valid.
\begin{table}
\begin{tabular}{l *{8}{| c}}
$n$ mod $8$ & 0 & 1 & 2 & 3 & 4 & 5 & 6 & 7\\
\hline
GS degeneracy & $\sqrt{1}$ & $\sqrt{2}$ & $\sqrt{4}$ & $\sqrt{8}$ & $\sqrt{4}$ & $\sqrt{8}$ & $\sqrt{4}$ & $\sqrt{2}$
\end{tabular}
\caption{Ground state degeneracy in the presence of generic 4-fermion interactions as a function of the total number of MZMs $n$ modulo 8. Irrational ground state degeneracy for odd numbers of MZMs implies that if we bring an additional MZM into the system but do not couple it to the system of interest, then the ground state degeneracy is multiplied by $\sqrt{2}$ and becomes rational. This is always a valid interpretation since in any physical systems the total number of MZMs is even.}
\label{tab:GS_degeneracy}
\end{table}

The $\mathbb{Z}_8$ periodicity is experimentally observable in a single electron tunneling spectroscopy. Similar to the non-interacting case we consider tunneling into the states of the $n$ Majoranas, either using scanning tunneling microscopy (STM) or a normal contact attached to the hole region. The tunneling can only take place if the matrix element of $\psi^\dag$ and $\psi$ between one of the ground states and the state into which we are trying to tunnel is non-zero. Otherwise one needs two particles to tunnel simultaneously, the process smaller in the tunneling amplitude, which we will neglect. We depict the positions of the lowest excited levels along with the possibility to observe them in tunneling conductance in Fig. \ref{fig:setups_conductance}(a).

\subsection{Array of vortices}

Another possible setup to observe the $\mathbb{Z}_8$ periodicity is depicted in the Fig. \ref{fig:setups_nMajoranas}(b). We consider a square array of single  vortices of size $2\times n/2$. For odd $n$ the last vortex is alone in the last line. In such a setup we need to make an assumption about the screening of the electron-electron interactions. As an example, we assume that all the vortices are closely situated, and the amplitude of the interaction between any 4 Majoranas is the same, $V_1$. This is an unnecessary assumption -- we have obtained qualitatively same results for different screening lengths -- but it removes multiple interaction strengths and simplifies the expressions.

The interactions, however, have dependence on the phase difference between the vortex cores.\cite{Chi15a}  This is the dependence we take into account to obtain the positions of energy levels of the system. These, and their respective  observabilities by single-electron tunneling are depicted in Fig. \ref{fig:setups_conductance}(b). We assume the possibility to tunnel into any of the vortices of the array, i.e. the tunneling probe much larger than the size of the array. Again, nothing qualitatively changes if we tunnel into a single generic vortex.

\subsection{General argument}

In this section we will discuss the ground state degeneracy for the system of $n$ interacting Majoranas. Due to locality of the edge states, this system is equivalent to an edge of a BDI wire with the quantum number $n$. We will use the following definition of the ground state degeneracy for a system of $n$ MZMs: it is the degeneracy of the ground state in the entire multi-dimensional parameter space of the Hamiltonian with all the interactions allowed by symmetry turned on. The degeneracy can be higher on a measure-zero regions of parameter space, where the accidental degeneracies are present on top the obligatory ones. The simplest example of such a region is the point where all the interactions are turned off. We assume there are no two regions of parameter space that are having non-zero measure and different ground state degeneracies. This assumption is justified by the zero-dimensional nature of the system, where there is no locality condition which can prevent two states from splitting.

It turns out that it is convenient to prove a more general statement: generically all the excited states in such systems will have the same degeneracy as the ground state. We will use the results of the more general argument and apply them to the ground state degeneracy. It is straightforward to expand the argument to all the excited states as well.

We can explicitly construct the states with the ground state degeneracies as shown in Table \ref{tab:GS_degeneracy} for $k=0\ldots 7$ MZMs. Examples of such constructions were just shown in the previous section. In all such examples the degeneracy of the ground and excited states was found the same as expected according to the general periodicity of Table \ref{tab:GS_degeneracy}. We now add to the group of $k$ Majoranas  $m$ blocks, each containing $8$ Majoranas with non-degenerate ground states. We thus obtain a system of $n = 8m + k$ Majoranas with the ground state degeneracy the same as for $k$ interacting Majoranas. Coupling these blocks together with a small enough coupling cannot \textit{increase} the degeneracy of the ground state, since one needs finite perturbation to close the gap. This limits the ground state degeneracy for $n$ Majorana bound states from above to the values in Table \ref{tab:GS_degeneracy}.

What is left is to prove that the above limit is actually the ground state degeneracy for $n$ Majoranas in a region of parameter space. For $k=0, 1, 7$ the limitation of the ground state degeneracy from below is obvious: for $k=0$ the degeneracy is $1$, while for $k=1$ and $7$ it is $\sqrt{2}$. The ground state degeneracy of at least $\sqrt{2}$ is always present for an odd number of Majoranas since then there is a single Majorana mode that can be decoupled from all the others, thus giving the desired $\sqrt{2}$ degeneracy.

We will prove the limitation on the ground state degeneracy from below by induction. The starting point is that for $n=0, 1, 2, 3$ no interactions term can be added to the Hamiltonian, thus the limitation of the ground state degeneracy from below is obvious. For $n=4$ there is a single interaction term allowed, as discussed above, it involves all the $4$ Majoranas. Such an interaction halves the ground state degeneracy, and makes a doubly-degenerate excited state \cite{Chi15b}. We now come to the step of induction, which will generalize the result to arbitrary $n$.

The induction step goes via \textit{reductio ad absurdum} path. Assume that we have  for $n = 8m + k$ Majoranas degeneracy lower than  expected. 

We already know that this is impossible for $k=0, 1, 7$. Let us proceed with the $k=2, 6$ case. The ground state degeneracy expected one reads off Table \ref{tab:GS_degeneracy} is $2$. Our assumption dictates that there is a region of the parameter space where the degeneracy is $1$ (and correspondingly the degeneracy of the excited states is $1$ as well). Now we use the fact that we have already proven our argument for the total number of Majoranas $n' = 8 - k$, where the ground state degeneracy is $2$, and the two ground states are different by a fermion particle number. 

Let us now bring the $n'$ Majoranas and $8m + k$ original ones together. The ground state of the decoupled system is one with an empty and an occupied fermion level. To couple the two ground states one needs a degenerate empty and an occupied fermion level in the rest of the system, which is absent due to the ground state degeneracy $1$. To prove this statement, one needs to rewrite the total Hamiltonian in the block-diagonal form, where one block corresponds to the empty fermion level, and another to the occupied. Then one notices that as each of the block is non-degenerate and as there is a symmetry of the total Hamiltonian under changing the occupation of the fermion, the doubly-degenerate ground state stays doubly-degenerate even after coupling to the rest of the system. Therefore for $8m + k + n' = 8(m+1)$ Majoranas there is a region of parameter space, where the ground state is doubly degenerate, which contradicts the assumptions, and consequently proves the statement for $k=2, 6$.

We can now add to $8m + 2$ Majoranas another two. We notice that there are two fermionic modes, empty and occupied, forming the ground state manifold without the coupling between the two systems. It is clear that since there is no direct tunneling between the Majoranas, the only coupling allowed in the model consists of the two new Majoranas and the occupation number of the old fermion. Therefore we are left with the ground state degeneracy $2$, but this degeneracy is between two states different by a bosonic operator, which proves the constraint on the ground state degeneracy for $k=4$.

Now we notice that $k=3, 5$ is different from $k=4$ by a single Majorana fermion. By adding or subtracting the Majorana it is impossible to remove the bosonic degeneracy of the ground state and the degeneracy should be enhanced by a factor $\sqrt{2}$ over $k=4$. This concludes the proof.

\section{Lapa-Teo-Hughes model}

\begin{figure}[tb]
\begin{center} 
\includegraphics[clip,width=1\columnwidth]{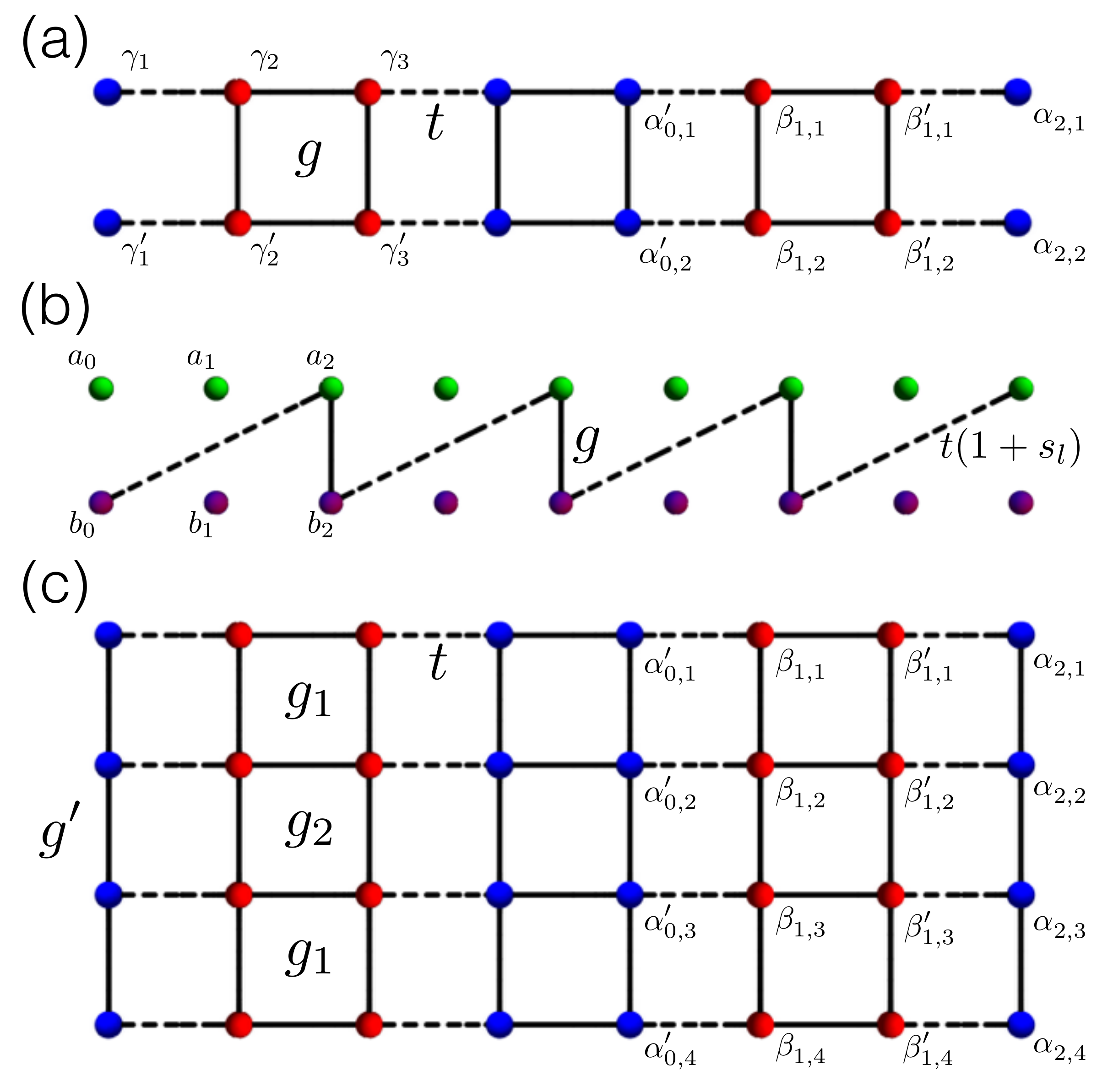}
\end{center}
\caption{(Color online) Lattice geometries for the interacting Majorana models. Color blue(red) indicates (anti)vortex Majorana modes, while solid and dashed lines indicate interaction and hopping terms, respectively. Panels a) and c) represent  the 2-leg and the  4-leg LTH ladder, respectively. These two models are invariant under particle-hole ($\Xi$), time reversal $\bar{\Theta}$, and inversion $\cP$ operations.  b) The 2-leg LTH model under the Majorana basis transformation, as discussed in the text.
}
\label{fig:leg_model}
\end{figure}

	We now turn to the 1d models of interacting MZMs. Lapa, Teo and Hughes\cite{Lap14} (LTH)  introduced a model of fermions in class BDI that is forced to be in the trivial phase by additional inversion symmetry in the absence of interactions. When interactions that respect the symmetries are introduced, the system, however, might transition to the topologically non-trivial phase characterized by gapped bulk and 4-fold ground state degeneracy associated with a pair of Kramers doublets bound to the two edges of the 1D system. An experimental realization of the 1D LTH model, using MZMs in vortices and antivortices in the surface of a TI, has been proposed\cite{Chi15b}. The proposed setup is depicted in Fig.\ \ref{fig:leg_model}(c) and consists of alternating clusters of 8 vortices and antivortices (that support 4-fermion interactions) connected to one another via tunneling terms. In the limit of strong interactions the ground state can be thought of as a direct product of unique ground states associated with each 8-vortex cluster. The ground state degeneracy is associated with the quartet of MZMs located at each end of the chain and is protected by $\bar{\Theta}$. This is a genuine interaction-enabled topological phase as no nontrivial phase can exist in a system with these symmetries in the absence of interactions\cite{Lap14}. Importantly, it persists as a stable phase for a finite strength of hopping $t$. 

In this Section we study the LTH model in detail by numerical techniques. We determine its phase diagram, ground state degeneracy as well as the entanglement spectrum. To facilitate this study we introduce a closely related 1D model that is exactly soluble. This is depicted in Fig.\ \ref{fig:leg_model}(a) and consists of alternating clusters of 4 vortices and antivortices. We call it the ``2-leg LTH model''. The original ``4-leg LTH'' model can thus be thought of as two coupled 2-leg models.
We also discuss potential realizations of the LTH-type models in 2D and 3D and their relevance to experimental systems. 

\subsection{2-leg LTH ladder}\label{2leg LTH}

The 2-leg LTH model Fig.\ \ref{fig:leg_model}(a) can be regarded as composed  of two parallel Majorana chains, which preserve time reversal $\bar{\Theta}$ and the reflection symmetry. 
Since the presence of inversion symmetry trivializes the Fidkowski-Kitaev ${\mathbb Z}_8$ 2-channel chain\cite{Lap14},  
this model should exhibit only topologically trivial phases even in the presence of interactions. Using the $\alpha$, $\beta$ notation as illustrated in the right half of Fig.\ \ref{fig:leg_model}(a) the Hamiltonian  is written as
\begin{align}
\cH=&-it \sum_{k=-M}^M \sum_{a=1}^2 (\alpha_{2k-1,a}'\beta_{2k,a}^{}+\beta_{2k,a}'\alpha_{2k+1,a}) \nonumber \\
&+g\sum_{k=-M}^M h'_{\Box}(\beta_{2k},\beta_{2k}')+g\sum_{k=-M}^{M-1} h'_{\Box}(\alpha_{2k+1}^{},\alpha_{2k+1}').
\end{align}
 We consider a system with $N'$ Majorana sites along the chain, $N'$ being a multiple of 4.  For convenience we define (half) integer valued $M=(N'-4)/8$. Index $k$ is then also (half) integer and extends between $-M$ and $M$. The interaction $h'_{\Box}(\lambda_n,\lambda'_n)=\lambda_{n,1}\lambda'_{n,1}\lambda_{n,2}\lambda'_{n,2}$. 
The hopping between the same type of Majoranas, which breaks the time reversal symmetry, is absent in the Hamiltonian. The symmetries allow for interaction 
terms containing two $\alpha$ and two $\beta$ operators but we expect these to be small compared to the direct hopping terms already included in $\cH$.
We thus include only 4-fermion terms between the same type of MZMs.

The model is integrable because it has an extensive number of constants of motion. Specifically, it is easy to see that products 
\begin{eqnarray}\label{cms}
\Lambda^{(1)}_k&=&\alpha'_{2k-1,1}\alpha'_{2k-1,2}\beta_{2k,1}\beta_{2k,2} \\ \Lambda^{(2)}_k&=&\beta'_{2k,1}\beta'_{2k,2}\alpha_{2k+1,1}\alpha_{2k+1,2} \nonumber
\end{eqnarray}
 commute with $\cH$ for all $k$. In the following we introduce a non-local transformation that allows one to replace these operators  by $c$-numbers and maps the problem onto a collection of non-interacting 1D Kitaev chains.

To effect the transformation it is expedient to relabel the MZM operators as indicated in the left half of Fig.\ \ref{fig:leg_model}(a).
The Hamiltonian can be written in an economical way
\begin{align}
\cH = &- i t \sum_{l=0}^N (\gamma_{2l+1}^{} \gamma_{2l+2}^{} + \gamma'_{2l+1}\gamma'_{2l+2}) \nonumber  \\
&- g \sum_{l=0}^{N-1}\gamma_{2l+2}^{} \gamma_{2l+3}^{} \gamma'_{2l+2}\gamma'_{2l+3}, \label{2leg simple}
\end{align}
where $N=(N'-2)/2$. 

The operators transform $\gamma_{j}\rightarrow (-1)^j\gamma'_{2N+3-j}$ and $\gamma'_{j}\rightarrow (-1)^j\gamma_{2N+3-j}$ under the inversion operation, of which the inversion center is located at the middle of the $(N+1)$-th and $(N+2)$-th sites since $(\alpha,\alpha',\beta,\beta')\rightarrow (\alpha', -\alpha, \beta', -\beta)$. (The minus signs are coming from the spin-$1\over 2$ of the Fu-Kane model, see \ref{setup LTH}.) Since inversion symmetry in a 1D system plays an effective role of reflection symmetry, it will be shown in \ref{setup LTH} that the Hamiltonian is invariant under all of the symmetries in class BDI$+\cR_{--}$.\cite{Chi15,chiu_reflection}

\begin{figure*}[tbh]
\begin{center} 
\includegraphics[clip,width=2\columnwidth]{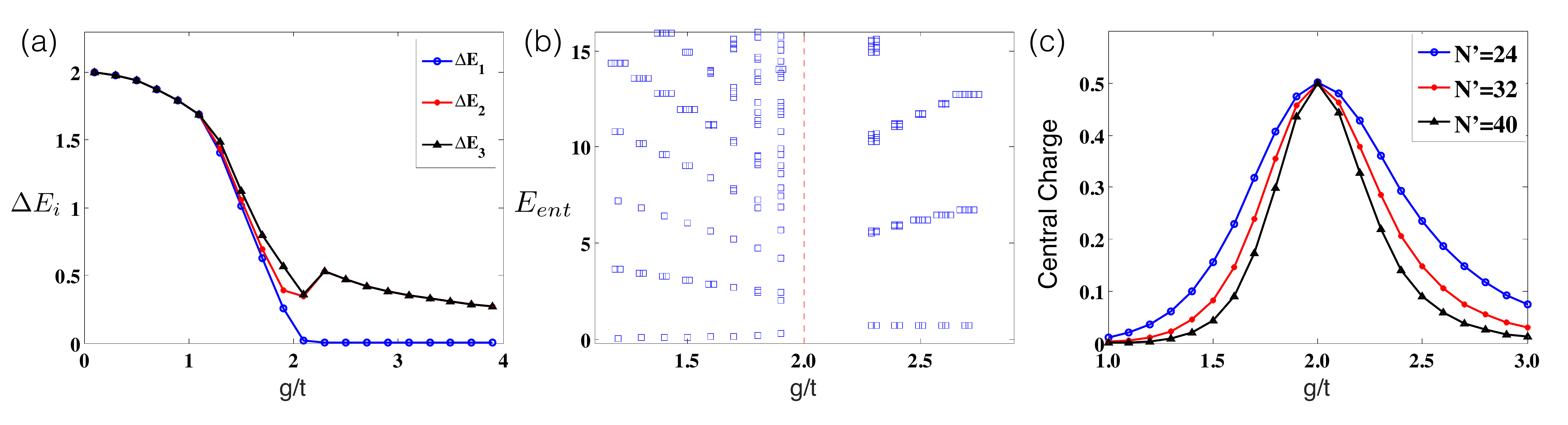}
\end{center}
\caption{(Color online) The numerical results of the 2-leg LTH model are computed using DMRG as $g/t$ varies. Here $N'$ indicates the length of the chain, $N'=2N+2$ in Eq.\ (\ref{2leg simple}). a) $\Delta E_i= E_i- E_0$ is the energy difference between the ground state and the $i$-th excited state in a system with open boundary conditions. b) Entanglement spectrum for periodic boundary conditions. c) The central charge $c$. At the critical point it is well defined and size-independent. Furthermore, the energy gap closes and  the number of the degenerate states in the entanglement spectrum changes at $g/t=2$. Thus, the critical point is at $g/t=2$ as expected based on the analytic solution. In panels (a) and (b), $N'=120$.
}
\label{2leg plot}
\end{figure*}	
Any 1D chain in this reflection symmetry class without interactions is always in the trivial phase. For this 2-leg model  even in the presence of the interactions the system is still trivial, 
but it is exactly solvable. 

The transformation, analogous to the Jordan-Wigner transformation is
\begin{align}\label{transf1}
\gamma_j =& P_j \bigg(
\prod_{k\;\mathrm{odd}}^{k<j} i b_k a_k\bigg) i b_{j-1} a_j, \\
\gamma'_j = &- i P_j \bigg(
\prod_{k\;\mathrm{even}}^{k<j} i b_k a_k\bigg) i b_{j-1} a_j, \\
\end{align}
where 
\begin{equation}
P_j =\left\{\begin{array}{c}
 1, \; j \; \mathrm{odd},\\
-i, \; j \; \mathrm{even}.
\end{array}\right.
\end{equation}
Here $a_k$ and $b_k$ are Majorana operators satisfying 
\begin{align}
\{a_k, a_{k'}\} = 2 \delta_{k, k'}, \; \{b_k, b_{k'}\} = 2 \delta_{k, k'}, \; \{a_k, b_{k'}\} = 0, 
\end{align}
and $k$ runs from $0$ to $N$. It is straightforward to check that this transformation does not change the original $\gamma$, $\gamma'$ commutation relations. Now we can rewrite the original Hamiltonian in the new Majorana basis. For odd $j$ it holds
\begin{align}
\gamma_j \gamma_{j+1} =& b_{j-1} a_{j+1}, \\
\gamma'_j \gamma'_{j+1}  =& i b_{j-1} a_j b_j a_{j+1},
\end{align}
while for even $j$
\begin{align}
\gamma_j \gamma_{j+1} = &-i b_{j-1} a_j b_j a_{j+1}, \\
\gamma'_j \gamma'_{j+1} = & - b_{j-1} a_{j+1}.
\end{align}
Therefore the Hamiltonian is rewritten as 
\begin{align}\label{hab}
\cH = &- it \sum_{l=0}^N b_{2l} a_{2l+2}(1 + i a_{2l+1}b_{2l+1})   \nonumber \\
&+ g \sum_{l=0}^{N-1}ia_{2l+2} b_{2l+2}.
\end{align}
We note that the number of Majoranas in $a,\ b$ basis is two more than the number of Majoranas in $\gamma,\ \gamma'$ basis as shown in Fig. \ref{fig:leg_model}(b). The reason is that we have introduced two extra free Majoranas $a_0$ and $b_{2N+2}$. These are necessary to satisfy the canonical commutation relations and  the redundancy should be kept in mind when discussing the ground state degeneracy in the transformed basis. 

After the transformation, it is clear the system has many integrals of motion $s_l = ia_{2l + 1} b_{2l+1}$. These are the same quantities as defined in Eq.\ (\ref{cms}). We can replace them by their respective eigenvalues $s_l=\pm 1$.
In each sector, labeled by the set of quantum numbers $\{s_l\}$, the transformed Hamiltonian is non-interacting and represents a ``broken'' Kitaev chain illustrated in Fig. \ref{fig:leg_model}(b). It has alternating bonds with hopping strength given by $g$ and $(1+s_l)t$. 
In the sector with all $s_l = +1$ this model describes a single Kitaev chain with a phase transition between its two gapped phases at $t = g/2$. 
In the sector with all $s_l = -1$ the system consists of disconnected monomers and the spectrum is a pair of flat bands; $t$ does not enter. In the mixed sector, where some hoppings are present and some are absent, we have an array of the disconnected Kitaev chains of finite length. It is clear that the wider the bandwidth of the Hamiltonian, the lower the energy of the ground state in a given sector. Therefore, we expect the absolute ground state to be in the $s_l=+1$ sector. We have confirmed this by explicit numerical simulation of the Hamiltonian (\ref{hab}) on a lattice up to $19$ integrals of motion ($20$ Majoranas in the hopping model, $58$ in total).

We now study the ground state $s_l=+1$ sector. It is known that the Kitaev chain is dual to the transverse-field Ising model.\cite{Fendley} The phase transition between its two gapped phases is therefore in the transverse-field Ising model universality class. Indeed it is possible to transform the 2-leg LTH Hamiltonian onto a set of broken Ising chains, either directly going from $\gamma$ operators to spin-$1\over 2$ operators, or in two steps via the broken Kitaev chain Eq.\ (\ref{hab}). This also helps to understand the nature of the gapped phases of the original model. In the transformed basis for $t> g/2$ the absence of ground state degeneracy leads to the topologically trivial phase. For $t<g/2$ Majorana zero modes appear at each end in the $a,\ b$ basis. When we assume $N\gg 1$, the form of the end modes is given by 
\begin{align}
B=&\frac{1}{1+2t/g}\sum_{l=0}^{N}\left(-\frac{2t}{g}\right)^l b_{2l},   \\
A=&\frac{1}{1+2t/g}\sum_{l=0}^{N}\left(-\frac{2t}{g}\right)^l a_{2N-2l}. 
\end{align}	
The	fermion operator $B+Ai$ switches one ground state to the other. Although these Majorana operators are localized in $a,\ b$ basis,  the fact that the transformation (\ref{transf1}) is non-local means that in the 
original $\gamma, \gamma'$ basis these operators do not represent edge degrees of freedom. In the original basis $t<g/2$ is a conventional broken symmetry phase, akin to the ferromagnetic phase in the Ising model. [The symmetry that is spontaneously broken in Hamiltonian (\ref{2leg simple}) for $t<g/2$ is generated by $\gamma_k'\to -\gamma_k'$. It remains unbroken in the other phase.]
The 2-leg LTH model undergoes a phase transition but both phases are topologically trivial as expected on the basis of general arguments presented above. 

To confirm the above conclusions we have performed DMRG calculations to compute the energy spectrum, the entanglement spectrum, and the central charge of the system as a function of dimensionless coupling $g/t$.  The spectra in panels  (a), (b) of Fig.\ \ref{2leg plot} show that the phase transition indeed occurs at $g/t=2$ as the gaps close. The ground state is doubly degenerate for $g>2t$ and the entanglement spectrum also shows 2-fold degeneracy in the symmetry broken phase. Importantly, the ground state degeneracy in this case occurs for both open and periodic boundary conditions, confirming that $g>2t$ is a conventional broken symmetry phase. To confirm the phase transition property, we numerically compute the central charge $c$ using its relation with the entanglement entropy of ground states,\cite{PCal09}
\begin{equation}
S(n)=\frac{c}{3}\log(\frac{N'}{\pi}\sin\frac{\pi n}{N'})+S_0.
\end{equation}
Here $N'$ is the system size, $n$ is the size of subsystem, $S(n)$ is the corresponding entanglement entropy and $S_0$ is a constant. At the critical point Fig.\ \ref{2leg plot}(c) indicates  the value $1\over 2$ of the central charge in agreement with our expectation that the phase transition is in the universality class of the transverse-field Ising model.

\subsection{4-leg LTH ladder}\label{4leg LTH}

\begin{figure*}[tb]
\begin{center} 
\includegraphics[clip,width=2\columnwidth]{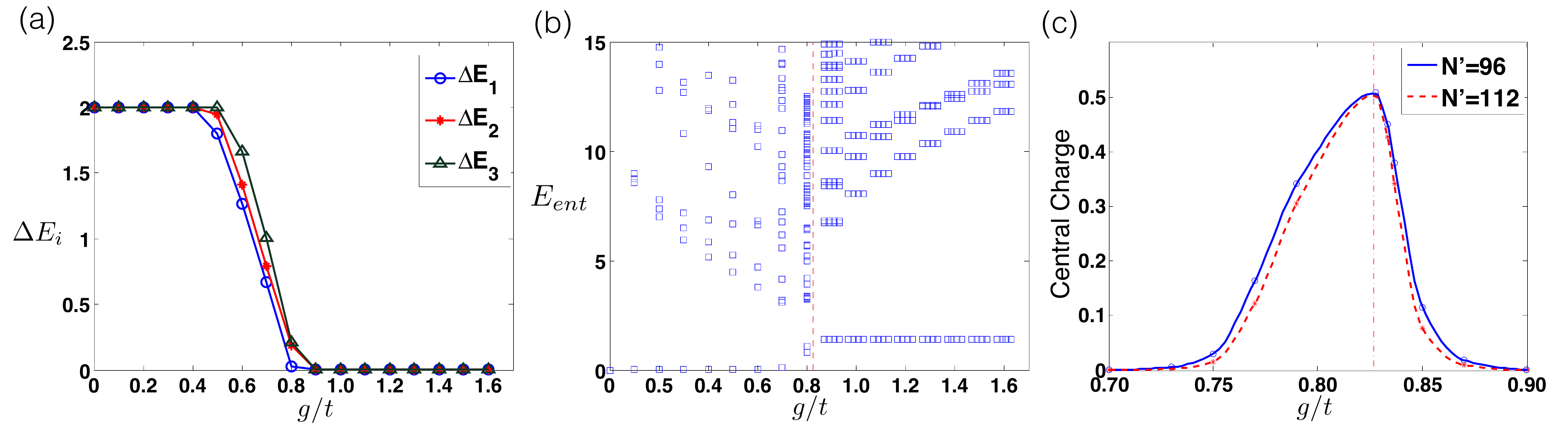}
\end{center}
\caption{(Color online) Numerical results for the 4-leg LTH model. We define the interaction strength $g\equiv g_1=g_2=g'$ and perform DMRG with the system size $N'=8M+4=120$. a) $\Delta E_i= E_i- E_0$ is the energy difference between the ground state and the $i$-th excited state for the 4-leg LTH with open boundary conditions. b) The entanglement spectrum of the ground state for the 4-leg LTH with periodic boundary conditions as a function of $g/t$. The LTH model exhibits 4-fold degeneracy for the ground state and the entanglement spectrum  when  $g/t$ is larger that the phase transition point  $g_c/t\approx 0.83$. c) The central charge as a function of $g/t$.}
\label{4leg plot}
\end{figure*}

We proceed with the examination of the 4-leg LTH model whose geometry is illustrated in Fig. \ref{fig:leg_model}(c). In the absence of interactions the restriction of inversion symmetry prohibits any topologically non-trivial state. Let us start with a non-interacting 4-leg setup to illustrate this. Due to the restriction of the symmetries, only hopping terms between $\alpha$ and $\beta$ MZMs are present in the Hamiltonian. If we focus on the nearest neighbor such hoppings the Hamiltonian reads  
\bee
\cH_{\rm{hop}}=-i\sum_{k=-M}^M \sum_{j=1}^4 t(\alpha_{2k-1,j}'\beta_{2k,j}^{}+\beta_{2k,j}'\alpha_{2k+1,j}),
\ee
where $M$ is a (half) integer. The system is formed by a set of Majorana dimers and, importantly, all the Majoranas in the system are  hybridized by the hopping.
Therefore, there are no edge states and the system is topologically trivial. To generate the topological phase, we introduce reflection symmetry preserving interactions. The dominant interaction involves groups of four Majoranas of the same type, 
\begin{align}
{h}_{|}({\bf \lambda}_l)=&g'\lambda_{l,1}\lambda_{l,2}\lambda_{l,3}\lambda_{l,4}, \\
h_{\Box}(\lambda_l^{},\lambda_l')=&g_1(\lambda_{l,1}^{}\lambda_{l,1}'\lambda_{l,2}^{}\lambda_{l,2}'+\lambda_{l,3}^{}\lambda_{l,3}'\lambda_{l,4}^{}\lambda_{l,4}') \nonumber \\
&+g_2\lambda_{l,2}^{}\lambda_{l,2}'\lambda_{l,3}^{}\lambda_{l,3}' 
\end{align}
where $\lambda_{l,j}$ stand for either $\alpha$ or $\beta$ MZM. 
As shown in Fig.\ \ref{fig:leg_model}(c) the Hamiltonian of the open chain including the interactions can be written as 
\begin{align}
\cH_{\rm{LTH}}=&\cH_{\rm{hop}}+\cH_{\rm{int}},  \label{4leg LTH eqn}
\end{align}
where
\begin{align}
\cH_{\rm{int}}=&\sum_{k=-M}^{M}\big [{h}_{|}(\alpha_{2k-1}')+{h}_{|}(\beta_{2k}^{})+{h}_{|}(\beta_{2k}')+{h}_{|}(\alpha_{2k+1}^{})\big] \nonumber \\
&+\sum_{k=-M}^M h_{\Box}(\beta_{2k},\beta_{2k}')+\sum_{k=-M}^{M-1} h_{\Box}(\alpha_{2k+1}^{},\alpha_{2k+1}').
\end{align}
To implement periodic boundary conditions, we add an extra interaction term $h_{\Box}(\alpha_{2M+1}^{},\alpha_{-2M-1}')$ to  the Hamiltonian. The Hamiltonian is invariant under inversion symmetry operation $\{\alpha_{l,j},\alpha_{l,j}'\}\rightarrow \{\alpha_{-l,5-j}',-\alpha_{-l,5-j}\} $ and $\{\beta_{l,j},\beta_{l,j}'\}\rightarrow \{\beta_{-l,5-j}',-\beta_{-l,5-j}\}$ since the Fu-Kane model is an intrinsic spin-$1\over 2$ system (see \ref{setup LTH}).

Let us first consider the extreme case when the hopping is off, $t=0$. The Hamiltonian then consists of decoupled clusters each containing 8 (anti)vortex Majoranas. We may compute the ground state of each such cluster described by
\bee
\cH_{\rm{sub}}={h}_{|}({\bf \lambda})+{h}_{|}({\bf \lambda}')+h_{\Box}(\lambda^{},\lambda')
\ee
Assume $g_1,\ g_2,\ g'$ are positive and define 4 complex fermions $d_{j,a}=(\lambda_{j,a}^{}+i\lambda_{j,a}')/2$ in each cluster. The many-body wavefunction can be expressed in the fermion basis $\ket{n_1n_2n_3n_4}$, where $n_a$ is the eigenvalue of the fermionic number operator ($\hat{n}_{j,a}=d_{j,a}^\dagger d_{j,a}$).   The unique ground state is then given by $(\ket{0000}-\ket{1111})/\sqrt{2}$  with energy $-2g_1-g_2-g'$.  As a result, with the periodic boundary conditions the 4-leg LTH model has a unique ground state in the limit when $t=0$. Since there is a gap to the lowest excited state we expect this non-degenerate ground state to survive for some range of non-zero $t$.

For open boundary conditions that respect the reflection symmetry, four $\alpha_{-2M-1,a}'$ at the left end and four $\alpha_{2M+1,a}$ at the right end have only interactions described by ${h}_{|}(\alpha_{-2M-1}')$ and ${h}_{|}(\alpha_{2M+1}^{})$ respectively. It can be easily seen that four many-body Majorana operators constructed from the edge MZMs 
\begin{align*}
\alpha_{-2M-1,1}'\alpha_{-2M-1,2}',&\ \ \  \alpha_{-2M-1,3}'\alpha_{-2M-1,4}',\\ 
\alpha_{2M+1,1}^{} \alpha_{2M+1,2}^{} ,&\ \ \ \alpha_{2M+1,3}^{}\alpha_{2M+1,4}^{}
\end{align*} 
commute with the full Hamiltonian. This implies 4-fold degeneracy of the ground state associated with the ends of the chain. As argued in Ref.\ \onlinecite{Lap14} the emergent time-reversal $\bar{\Theta}$ acts anomalously in this degenerate subspace (such that $\bar{\Theta}^2=-1$) and the states can thus be viewed as two Kramers doublets. We therefore expect the degeneracy to be robust against any perturbations that do not break $\bar{\Theta}$ and as long as the bulk remains gapped.

We now study the general case with hopping $t$ turned on by means of DMRG. For simplicity and concreteness we take $g\equiv g_1=g_2=g'$ but this is by no means essential. Fig.\ \ref{4leg plot}(a) shows 4 lowest energy states obtained by DMRG as a function of $g/t$ for open boundary conditions. This indicates a phase transition at $g_c/t\approx 0.83$ to the state with a 4-fold degenerate ground state, in agreement with the above analysis. Importantly, no such  degeneracy is observed for periodic boundary conditions which confirms the topological character of the interacting phase.

An alternative way to confirm the non-triviality of the phase is to consider the degeneracy of the entanglement spectrum. To compute the entanglement spectrum, the LTH model with periodic boundary condition is separated by the two vertical cuts between horizontally closest (anti)vortices. In agreement with the ground state degeneracy of the open chain, the entanglement spectrum Fig.\ \ref{4leg plot}(b) exhibits four-fold degeneracy since each cut contributes two-fold degeneracy from the wavefunction $(\ket{0000}-\ket{1111})/\sqrt{2}$ at the cut.
  
Fig.\ \ref{4leg plot}(c) displays the central charge computed using DMRG as a function of $g/t$. Similar to the case of the 2-leg LTH model it saturates at a size-independent value of $c={1\over 2}$ when $g=g_c$. This suggests that the phase transition in the 4-leg LTH model is in the transverse-field Ising universality class. Understanding more fully the microscopic origin of this transition is an interesting problem which we leave for future study.

	For topological crystalline insulators and superconductors, the bulk topology is commonly determined by 
	the presence or absence of protected gapless modes  at the boundary, that are invariant under a spatial symmetry. However, this is not the case for this interaction-enabled topological phase. Each end of the LTH model, which maps onto the other under inversion, is not invariant under inversion so the inversion symmetry does not protect the Majorana end-modes. Instead, the LTH model inherits its topology from the class BDI in 1D, which hosts stable Majorana boundary modes protected by time reversal symmetry and particle-hole symmetry. The presence of these Majorana boundary modes can be induced by symmetry-preserving interactions even if the phase would otherwise be trivial in the non-interacting system with the same symmetries.

\subsection{LTH-type  models in dimensions 2 and 3}

In two dimensions the interaction-enabled phases are possible in symmetry class BDI with an additional $C_4$ rotation symmetry. In the non-interacting case with weak invariants $\nu_x$ and $\nu_y$ in $x$ and $y$ directions respectively from 1D class BDI, the symmetry requires $\nu_x = \nu_y$ and $\nu_y = -\nu_x$. The only solution is $\nu_x=\nu_y=0$ and all such systems must be topologically trivial. In the interacting case, as before, the 1D classification changes to ${\mathbb Z}_8$ and $\nu_x = \nu_y = 4$ becomes another possible solution of the above criterion indicating an interaction-enabled topological phase. A similar argument can be made for a 3D system with  $C_4$ rotation symmetry about three orthogonal axes.
\begin{figure}[t]
\begin{center} 
\includegraphics[clip,width=8cm]{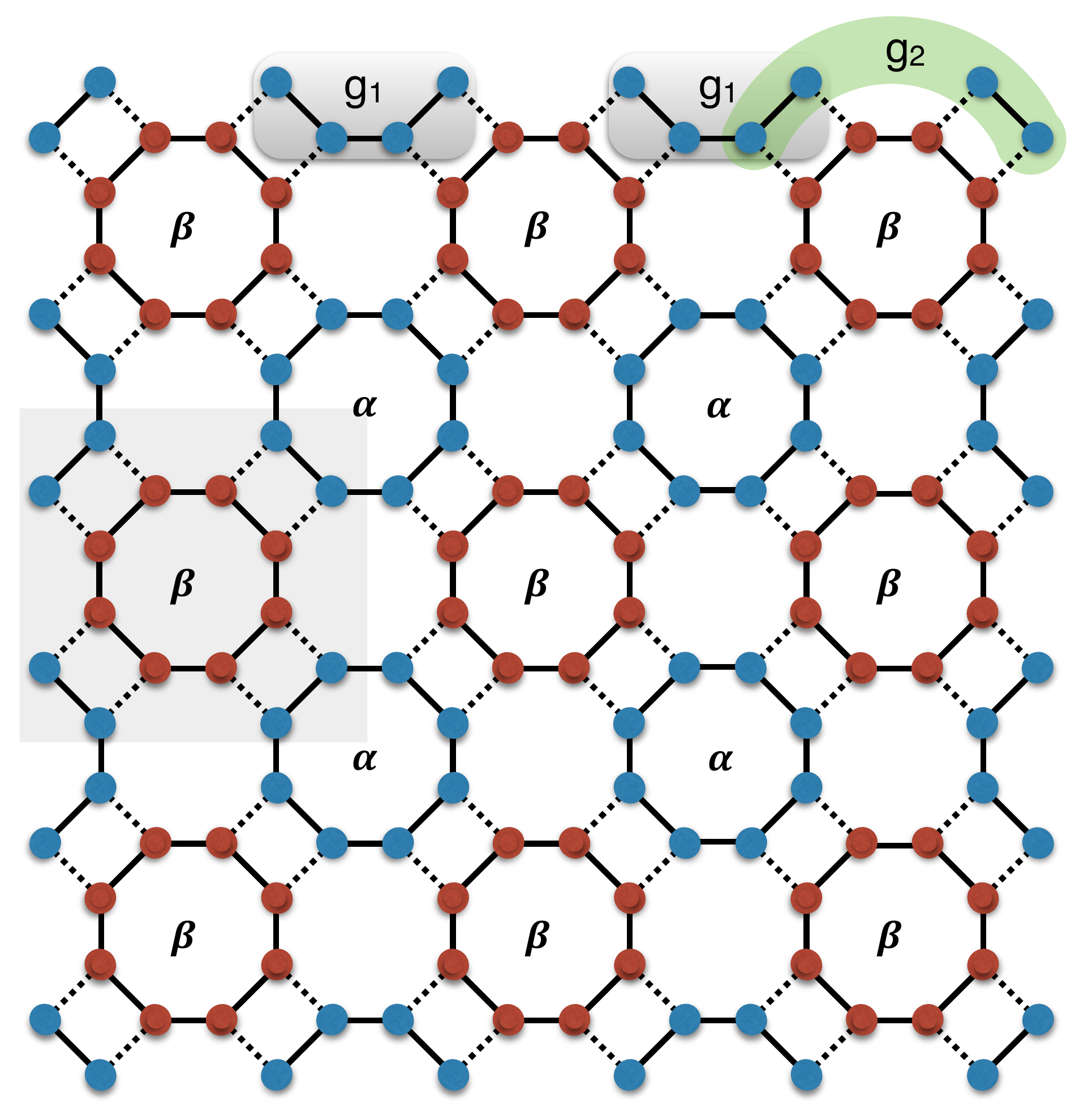}
\end{center}
\caption{(Color online) Proposed minimal geometry for an LTH-type model in 2D. Red (blue) circles indicate MZMs of type $\alpha$ ($\beta$). When the interaction dominates, the Majorana system in the bulk is gapped. The edge states are either gapless or spontaneously symmetry broken. }
\label{fig6}
\end{figure}

The 2D model we propose is depicted in Fig.\ \ref{fig6}. Similar to the 1D case it consists of alternating  clusters of 8 interacting MZMs connected to one another by hoppings. To understand the phase diagram of the model we notice that in the absence of interactions all the MZMs have counterparts connected by hoppings $t$, and therefore the ground state is unique with a gap $2t$ to all excitations.  In particular, the edge is also gapped. On the other hand, in the absence of hoppings and in the presence of generic interactions within each cluster containing 8 Majoranas the system is gapped in the bulk, as all the clusters have non-degenerate gapped ground state. At the same time we notice that the system has gapless flat-band edge states associated with the decoupled quartets of MZMs that exists at the boundary. Each cluster of 4 Majoranas on the edge has doubly-degenerate ground state. These two ground states are different by two Majorana operators, and therefore are related by emergent time-reversal symmetry $\bar{\Theta}$ that acts anomalously in this subspace\cite{Lap14} such that $\bar{\Theta}^2=-1$. Thus Kramers theorem applies and the edge state must either remain gapless or spontaneously break the time-reversal symmetry even away from the strongly interacting limit, as long as the bulk remains gapped.  In the latter case it is doubly-degenerate. Indeed we notice that if the hoppings are turned on, the edge spectrum  becomes gapped with a doubly degenerate ground state. This can be seen as follows. The hoppings will perturbatively generate additional interactions between the neighboring quartets of MZMs along the edge, such as that denoted in Fig.\ \ref{fig6} as $g_2$.  Effectively, then, the edge is described by a 1D interacting model discussed in Ref.\ \onlinecite{Rah15a}.  In the thermodynamic limit the model has been shown to exhibit a gapped doubly-degenerate ground state. 

\begin{figure}[t]
\begin{center} 
\includegraphics[clip,width=8cm]{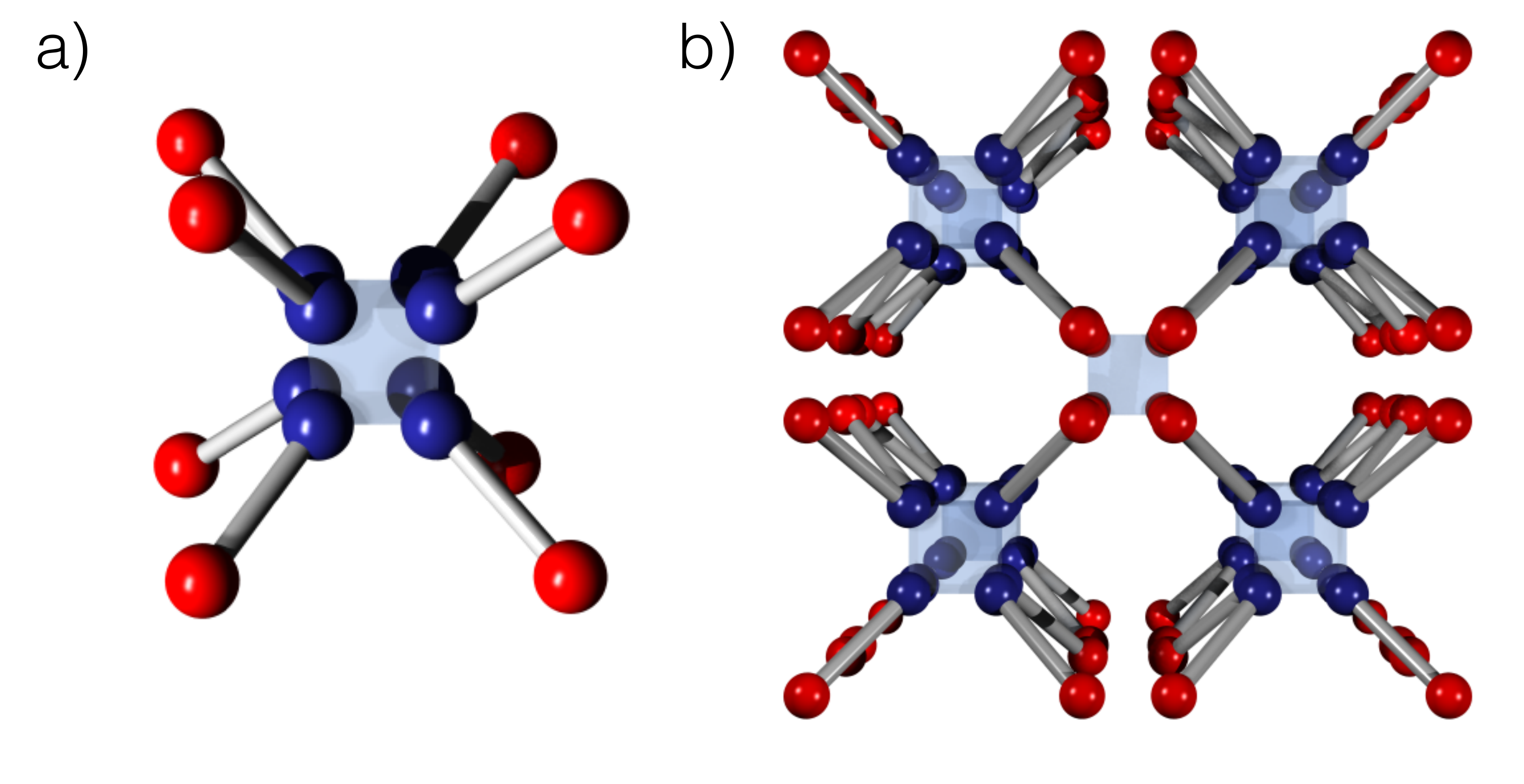}
\end{center}
\caption{(Color online) Proposed minimal geometry for an LTH-type model in 3D. Red (blue) spheres indicate MZMs of type $\alpha$ ($\beta$). a) The unit containing 8 MZMs of both types. b) The full 3D structure. Notice that surfaces consist of purely $\alpha$-type MZMs.
 }
\label{fig8}
\end{figure}

A similar construction can be given for a 3D system. A unit cell, containing 8 MZMs of each type is displayed in Fig.\ \ref{fig8}(a). When such unit cells are stacked to form a cubic lattice a 3D version of LTH-type model emerges, Fig.\ \ref{fig8}(b). The discussion of its phases parallels the discussion of the 2D case and we will not repeat it here. On that same basis we expect the model to exhibit two phases, one topologically trivial for weak interactions and one topological when interactions are strong. In the latter case the bulk once again is non-degenerate and gapped while the surfaces  perpendicular to the Cartesian axes are anomalous in that they exhibit either gapless states protected by  $\bar{\Theta}$ or spontaneously break $\bar{\Theta}$ and are then doubly degenerate and gapped.  


\subsection{Proposed experimental realizations of LTH-type models }\label{setup LTH}

\begin{figure*}[t]
\begin{center} 
\includegraphics[clip,width=1.6\columnwidth]{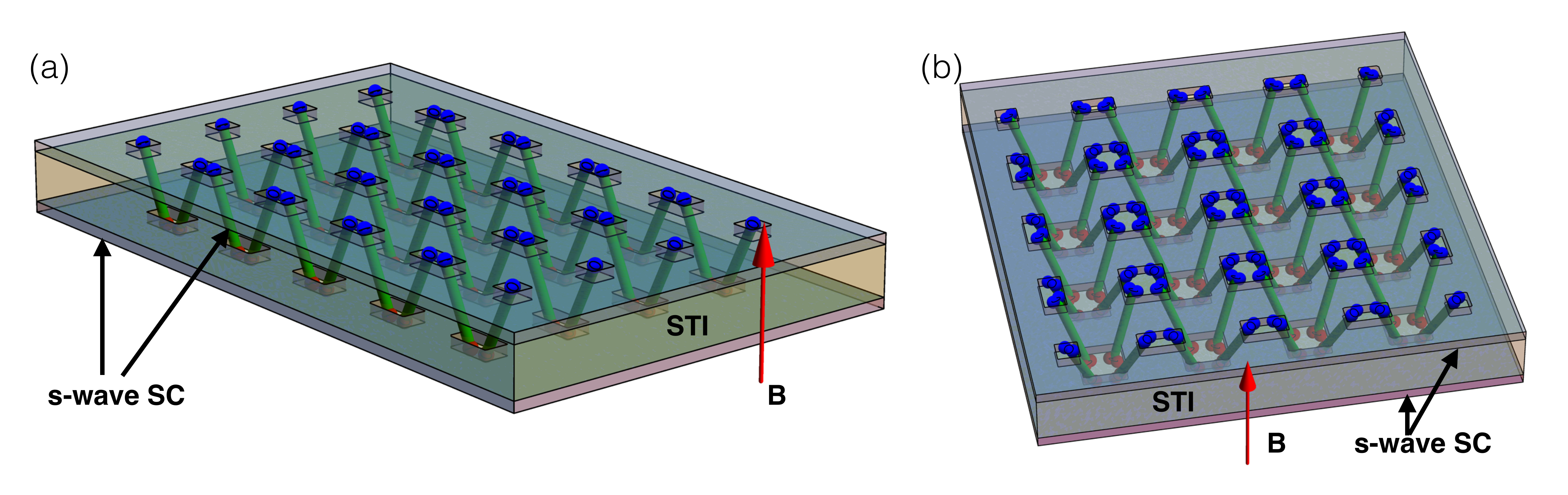}
\end{center}
\caption{(Color online) Proposed physical realizations of the of the LTH-type models in 1d and 2D. Superconductivity is induced in two surfaces of a thin TI film. The SC coating is patterned to create an array of pinning sites for vortices which then form structures that approximate the model geometries shown in Figs. \ref{fig6} and \ref{fig:leg_model}.
a) A quasi-1D system respecting the inversion symmetry can be used to realize the 4-leg LTH model. b) A 2D system with $C_4$ rotation symmetry that can be used to realized the LTH model in 2D. }
\label{fig7}
\end{figure*}
We now discuss potential realizations of the interaction-enabled phase for LTH-type models in dimensions 1 and 2. Our starting point is a thin film (or a flake) of an STI whose surfaces have been made superconducting by coating with a thin layer of an ordinary $s$-wave SC, such as Al or Nb. When a perpendicular magnetic field ${\bf B}$ is applied to this structure vortices are induced in the top surface while antivortices are induced in the bottom surface. MZMs bound to these are our basic ingredients. We now imagine that both surfaces are patterned with an array of holes in the SC layer as indicated in Fig.\  \ref{fig7}. The holes serve two purposes: (i) they pin vortices in the desired positions and (ii) they remove the undesired low energy Caroli-Matricon-de Gennes states that would otherwise exist in the cores of vortices in the ordinary SC. This way, vortex cores will only exist in the surface state of the TI and will be located in the right spatial positions. When the chemical potential $\mu$ of the TI is tuned to the Dirac point and when the TI film is sufficiently thin then the hopping amplitude $t$ will be appreciable only between MZMs located in the adjacent vortices and antivortices, indicated by green lines in  
Fig.\  \ref{fig7}. MZMs in the same surface are then connected only through 4-fermion interactions. If we furthermore take into account only the dominant interaction terms that occur between the quartets of MZMs that are closest together then it is easy to see how devices in Fig.\  \ref{fig6} approximate the LTH-type models in 1D (panel a) and 2D (panel b). We note that by layering the structure in  Fig.\  \ref{fig7}(b) it is possible to envision creating also the 3D model but in that case it is not clear how the pattern of hole might be fabricated to pin the vortices in the interior layers.

We now show that the LTH models in this heterostructure belong to reflection symmetry class BDI$+\cR_{--}$\cite{chiu_reflection} and a pair of (anti)vortices preserves inversion symmetry, which is equivalent to reflection symmetry in 1D. Consider two vortices located at positions $(\pm b, 0)$. The vortices we take into account by specifying a position-dependent phase of the pairing function
\bee
\Delta=\Delta_0 e^{i(\theta_++ \theta_-)},
\ee
where $\theta_\pm=\tan^{-1} \frac{y}{x \mp b}$. The pairing function is invariant under inversion $(x,y) \rightarrow (-x,-y)$ while the 2D reflection symmetry in any direction is broken.
With $\Delta(\bR)=\Delta(-\bR)$, the Hamiltonian $H_{\rm{FK}}$ in Eq.\ (\ref{eq:Hamiltonian}) is invariant under inversion operation 
\bee
\cP^{-1} H_{\rm{FK}} (-{\bf p},-{\bf R})\cP= H_{\rm{FK}} ({\bf p},{\bf R}),
\ee
where $\cP=i\tau^0\sigma^z$. Hence, we are able to construct 1D LTH models with the distribution of vortices and antivortices obeying $\Delta(\bR)=\Delta(-\bR)$ in Fig. \ref{fig7} (a).  
To check the topological classification, let the reflection symmetry operator be Hermitian $\cR=\sigma^z$ and  $\cR$ anticommute with $\bar{\Theta}$ and $\Xi$. Refs.\ \onlinecite{chiu_reflection,Lap14} show that the topology of such a 1D non-interacting chain in class BDI$+\cR_{--}$ is always trivial. 

We denote a pair of two Majorana zero modes located respectively at two (anti)vortices $(\pm b, 0)$ as $\alpha_b,\ \alpha_{-b}'(\beta_b,\ \beta_{-b}')$ in \ref{vortex antivortex}. They transform to $\alpha_{-b}',\ -\alpha_{b}(\beta_{-b}',\ -\beta_{b})$ under reflection symmetry since in a spin-${1\over 2}$ system $\cP^2=-1$. 

Similarly, the setup of 2D LTH model can be designed as shown in Fig.\ \ref{fig7} (b). The only difference is that each dot now represents two MZMs. This can be arranged by designing two pinning holes to be placed close to one another or else following more closely the octagonal pattern design displayed schematically in Fig.\ \ref{fig6}. In any case the holes are arranged so that the $C_4$ rotation symmetry is preserved. Furthermore, the symmetry has to be checked microscopically. 
Consider four vortices located at $(\pm b, 0)$ and $(0,\pm b)$. The phase of the pairing function coming from the four vortices is given by 
\bee
\Delta=\Delta_0 e^{i(\theta_{d,0}+\theta_{-d,0}+\theta_{0,d}+\theta_{0,-d})}
\ee
where $\theta_{x_0,y_0}=\tan^{-1} \frac{y-y_0}{x-x_0}$. It is easy to check that the pairing function is invariant under $C_4$ rotation symmetry operation $(x,y) \rightarrow (y,-x)$. In addition, the Fu-Kane Hamiltonian is invariant under $C_4$
\begin{align}
C_4^{-1}H_{\rm{FK}}\big ( (y,-x),(p_y,-p_x)\big ) C_4=H_{\rm{FK}}\big ( (x,y),(p_x,p_y)\big ).
\end{align}
where $C_4=\frac{1}{\sqrt{2}}(\bI +i \sigma_z)$. Hence, the design indicated in Fig.\  \ref{fig7}(b) obeys the $C_4$ symmetry and provides a possible realization the 2D LTH model.

\section{Discussion and conclusion}

Majorana zero modes bound to vortices in the SC surface of a topological insulator present a unique opportunity to study the effect of strong interactions in a fermionic system. This is because the kinetic energy of such fermions may be made to vanish by tuning a single parameter, chemical potential $\mu$ of the underlying TI. The interactions between MZMs, even if nominally not very strong, become the dominant energy scale in the problem. 
We have suggested a specific setup to experimentally probe the effect of  MZM interactions on the system's ground and excited states. We have discussed how to model 0D and 1D phases in this setup, and have proposed extensions to higher dimensions. Below we touch on the immediate experimental relevance of the setup discussed. 

The ingredients for the proposed setups are all in place. Superconductivity has been successfully induced  in the surface state of the 3D TI  by multiple groups.\cite{Kor11, Sac11, FQu12, Wil12, Cho13, YXu14, Zha14, PXu14} The ability to tune the  chemical potential  to the Dirac point as we require, has also been demonstrated.\cite{Cho13, YXu14, Zha14} Finally, vortices have been imaged in such devices\cite{PXu14} and spectroscopic evidence for MZMs in the cores of such vortices has been reported.\cite{xu2} We note that in the latter experiment the chemical potential was far away from the Dirac point so this specific system would not work for the purpose we envision in this paper. However, given the rapid pace of the experimental progress, we expect this last obstacle to be overcome in the near future.

Once the above hurdle has been surmounted  not much additional fabrication is required for our 0D Corbino ring proposal discussed in Sec.\ III. We envision testing the effect of interactions and the resulting Fidkowski-Kitaev ${\mathbb Z}_8$ periodicity before attempting to engineer the more complex 1D and 2D structures. The 0D system should provide a definitive signature of the presence of the MZM interactions by changing the periodicity of the occurrence of the zero-bias peak in the tunneling conductance. The energy scales of few degrees Kelvin we estimated show that the observation is possible with the currently available technology.

An ambitious, longer-term goal will be to engineer the 1D and 2D LTH-type models perhaps based on the designs outlined in Fig. \ref{fig7} using the same technology. We note that  besides creating the desired vortex patterns through artificial pinning, it is possible to obtain them naturally as a Josephson vortex lattice in a junction between two superconductors on top of a TI.\cite{Pot13} Probing the electron degrees of freedom in such systems is feasible using tunneling spectroscopy as has been demonstrated recently.\cite{Pill10,Seu08}

\section{Acknowledgment}

The authors thank I. Affleck, T. Liu, A. Rahmani and J. Teo  for useful discussions and correspondence.
The authors are indebted to  NSERC, CIfAR and Max Planck - UBC Centre for Quantum Materials for support. M.F.\ acknowledges The Aspen Center for Physics and IQMI at Caltech for hospitality during various stages of this project.


\end{document}